\documentclass[conference]{IEEEtran}
\IEEEoverridecommandlockouts
\usepackage[utf8]{inputenc}
\usepackage{graphicx}
\usepackage{xcolor}
\usepackage{amsmath}
\usepackage{multirow}

\begin{document}

\title{Periodic Online Testing for \\ Sparse Systolic Tensor Arrays}
\author{
\IEEEauthorblockN{Christodoulos Peltekis}
\IEEEauthorblockA{\small Electrical and Computer Engineering
\thanks{This work was supported by a research grant from Siemens EDA to Democritus University of Thrace for ``High Level Synthesis Research for Systems-on-Chip''}
\\ 
Democritus University of Thrace, Greece}
\and
\IEEEauthorblockN{Chrysostomos Nicopoulos}
\IEEEauthorblockA{\small Electrical and Computer Engineering\\  University of Cyprus, Cyprus}
\and
\IEEEauthorblockN{Giorgos Dimitrakopoulos}
\IEEEauthorblockA{\small Electrical and Computer Engineering\\  Democritus University of Thrace, Greece}}

\maketitle

\begin{abstract}
Modern Machine Learning (ML) applications often benefit from structured sparsity, a technique that efficiently reduces model complexity and simplifies handling of sparse data in hardware. Sparse systolic tensor arrays -- specifically designed to accelerate these structured-sparse ML models -- play a pivotal role in enabling efficient computations. As ML is increasingly integrated into safety-critical systems, it is of paramount importance to ensure the \textit{reliability} of these systems. This paper introduces an online error-checking technique capable of detecting and locating permanent faults within sparse systolic tensor arrays before computation begins. The new technique relies on merely four test vectors and exploits the weight values already loaded within the systolic array to comprehensively test the system. Fault-injection campaigns within the gate-level netlist, while executing three well-established Convolutional Neural Networks (CNN), validate the efficiency of the proposed approach, which is shown to achieve very high fault coverage, while incurring minimal performance and area overheads.
\end{abstract}


\section{Introduction}

Machine Learning (ML) techniques have enjoyed widespread proliferation in many different fields and are rapidly becoming ubiquitous. However, as ML models grow in size and complexity, they demand significant computational power and memory, making deployment on resource-constrained devices challenging. Model sparsification~\cite{hoefler2021sparsity} addresses this issue by reducing the number of weights, keeping only essential non-zero parameters. This approach enhances efficiency, reduces memory footprint, accelerates inference, and lowers energy consumption. These benefits make ML models more practical for real-time applications and deployment on edge devices.

\begin{figure}[t]
    \centering
    \includegraphics[width=0.98\columnwidth]{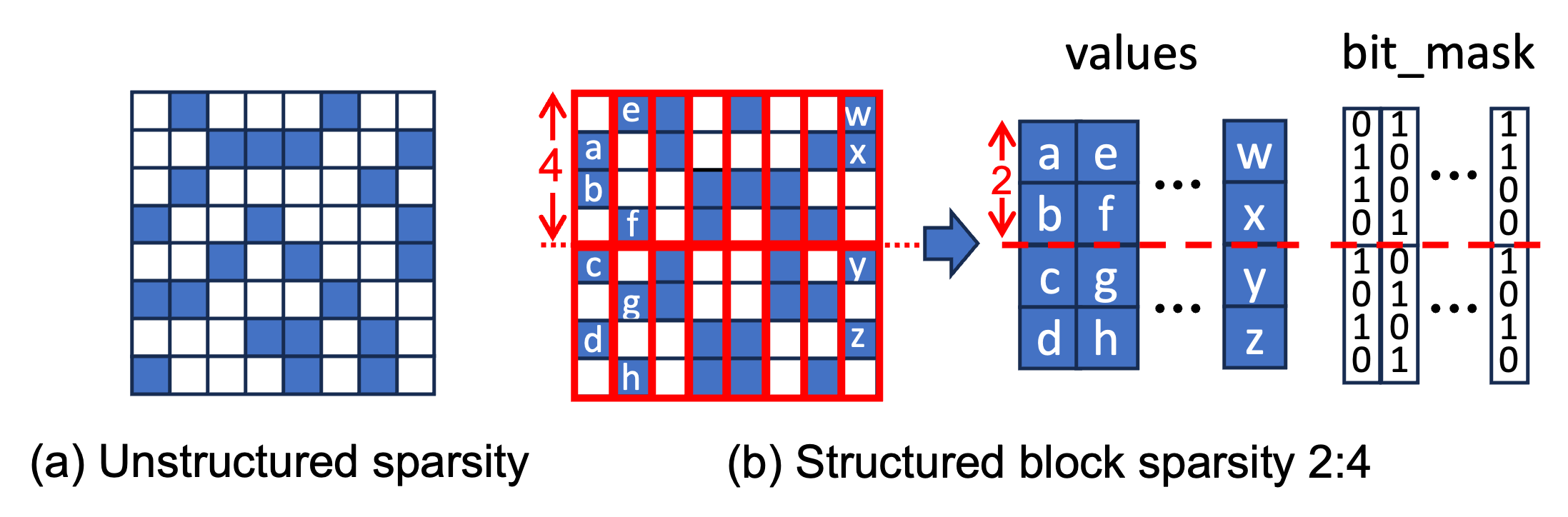}
    \caption{Example of (a) unstructured sparsity; and (b) structured block sparsity of 2:4 (i.e., up to 2 non-zero elements in every 4 consecutive elements in each column) and their respective packed storage with their associated bit masks.}
    \label{f:sparsity}
\end{figure}

There are two main types of sparsity. Under \textit{unstructured} sparsity~\cite{rigl}, there are no constraints on the distribution of the non-zero elements in the input, as abstractly illustrated in~Fig.~\ref{f:sparsity}(a). Consequently, extensive meta-data and multiple indexes are required to encode the non-zero locations. On the contrary, \textit{structured} sparsity~\cite{nvidia-block-sparse, learning-n-m} alleviates the meta-data cost and complexity by enforcing a constraint on the maximum possible number of non-zero elements present in every fixed-size block of consecutive input elements. Structured sparsity is typically defined by the notation $N$:$M$, which indicates that in each block of $M$ consecutive elements, at most $N$ may be non-zero. Fig.~\ref{f:sparsity}(b) shows an example of a $2$:$4$ structured sparse input matrix. The positions of the non-zero elements within each block is decided by the stored 4-bit masks. To accelerate directly in hardware the computations of structured-sparse data, sparse systolic arrays have been proposed, which take full advantage of the unique characteristics of structured block sparsity~\cite{sta, vegeta, s2ta, relaxed-structured}.

The pervasive adoption of ML in safety-critical application domains -- e.g., automotive~\cite{standard}, medicine~\cite{medical-cnn}, aviation~\cite{aviation-cnn}, etc. -- elevates the importance of the \textit{reliability} and/or \textit{fault-tolerance} attributes of the employed hardware accelerators. Typically, fault-tolerant systems rely on some form of fault-\textit{detection} technique/capability that triggers reactive action(s) upon fault detection. Periodic \textit{online testing} techniques allow for expedited in-the-field fault detection and can be very effective in rapidly detecting faults in mission-critical systems that cannot tolerate  delayed or offline fault detection. 

This paper proposes a novel periodic online self-testing methodology to detect permanent faults in sparse systolic arrays \textit{prior} to the commencement of the actual calculations by the running application.  Hence, a possible fault is identified \textit{before} any erroneous (thus, wasteful) work is performed. 
The proposed technique exploits the presence of already-stored data within the array to \textit{minimize} the number of required test vectors. Faults that occur during the computation can also be detected by Algorithm-Based Fault Tolerance (ABFT) techniques~\cite{abft, fault-tolerance-sa, abft-pratical, conv-checksum-tvlsi}, which \textit{concurrently} perform checks alongside the application computations.  
Under ABFT, faults are detected \textit{after} a computation is completed.

The proposed technique is inspired by the work in~\cite{runsafer}, which explores online self-testing in \textit{dense} systolic array architectures and reuses the stationary weights already loaded into the systolic array for testing purposes. In a similar vein, the technique proposed here also reuses the existing weight values, but it targets \textit{sparse} systolic arrays with their distinctive Tensor Processing Elements (TPE) and attributes that necessitate a more complex testing procedure, as compared to~\cite{runsafer}. 

The contribution of this work can be summarized as follows:
\begin{itemize}
\item
By utilizing the weight values of the running application during the testing phase, the presented technique requires merely {\bf\textit{four} test vectors} to provide very high coverage against permanent faults, which minimizes the time lost for on-line testing before the initiation of computation.
\item
The proposed approach reuses the systolic array's existing storage elements to minimize the incurred hardware overhead, as compared to other self-testing techniques that use scan chains~\cite{strait}.
\item
To assess the achieved fault coverage of the proposed technique, we employ random fault injections at the gate level, while executing three established Convolutional Neural Networks (CNN) applications. The obtained results demonstrate very high achieved coverage across all three benchmarks, thereby validating the effectiveness of the new lightweight online checking mechanism. Additionally, the incurred overheads to the system performance and the hardware area are shown to be minimal.
\end{itemize}

\section{Background: Sparse Systolic Tensor Arrays}
\label{s:background}

\begin{figure}
    \centering
    \includegraphics[width=1 \columnwidth]{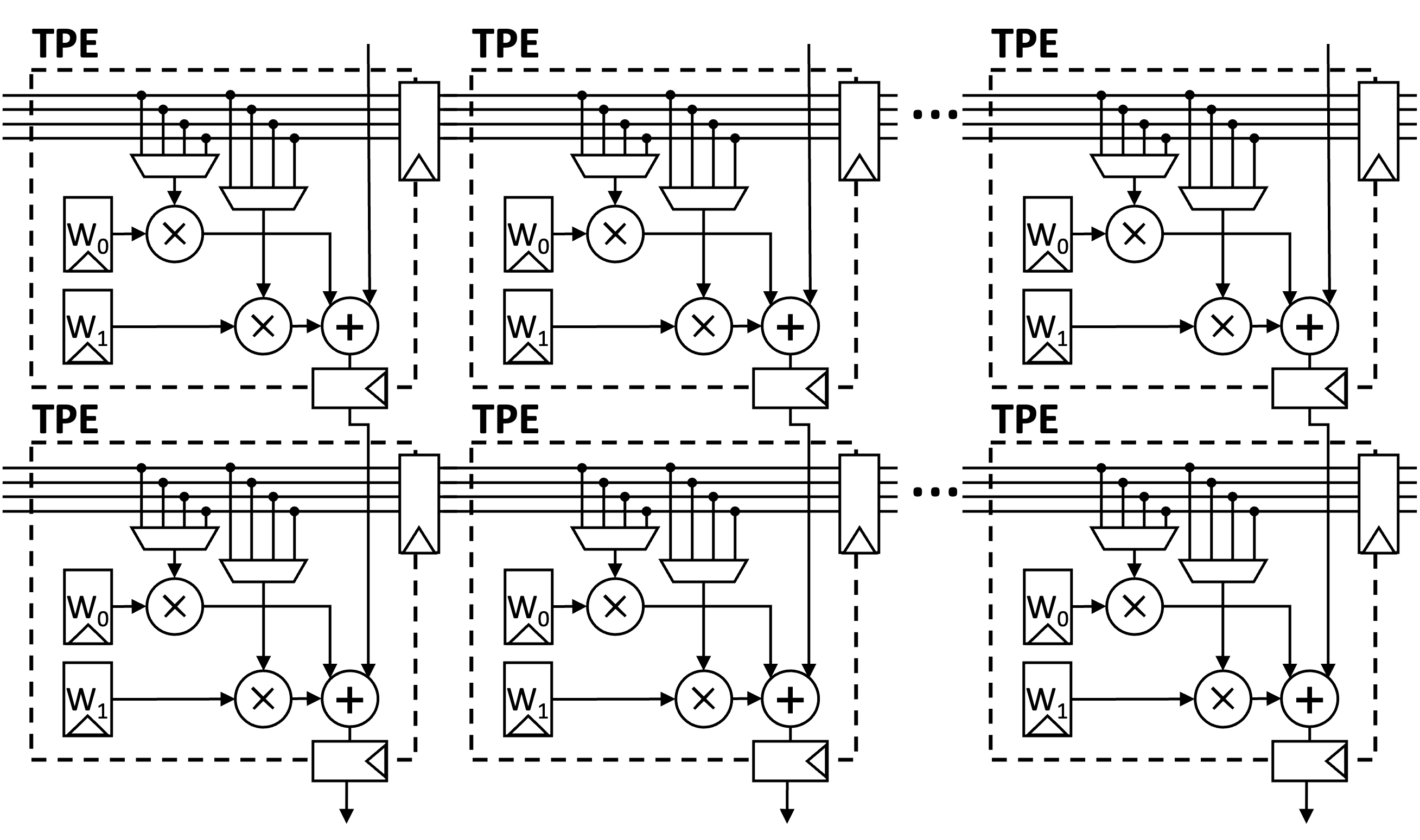}
    \caption{A sparse systolic tensor array that employs the weight-stationary dataflow; i.e., the weights are pre-loaded into the Tensor Processing Elements (TPE) and remain stationary during the operations. The inputs and outputs flow in the horizontal (west-to-east) and vertical (north-to-south) directions, respectively.}
    \label{f:sparse-tensor-array}
\end{figure}

A sparse systolic tensor array computes the matrix product $C$=$A\, W$ of a dense input matrix $A$ and a structured-sparse weight matrix $W$. It adopts the same systolic array architecture as a conventional \textit{dense} array, but, instead of using a \textit{scalar} Processing Element (PE) as its building block, it employs a more complex \textit{Tensor} PE, aptly called TPE~\cite{sta}. Scalar PEs in a Weight-Stationary (WS) dataflow accept a single input value and use a single local (pre-loaded) weight value to perform a Multiply-Accumulate (MAC) operation per cycle. In contrast, Tensor PEs take a \textit{block} of $M$ consecutive input values and use up to $N$ local weight values to perform up to $N$ multiplications in each MAC operation. This micro-architecture supports $N$:$M$ structured block sparsity. The architecture of a two-row sparse tensor array is illustrated in Fig.~\ref{f:sparse-tensor-array}. This array is configurable and can support both $2$:$4$ and $1$:$4$ structured sparsities. 

As shown in the figure, each TPE consists of two weight registers storing the (up to) two non-zero weight elements identified within every four elements of the pruned weight matrix. These weight registers are loaded with the appropriate weights during a distinct weight-loading phase, as dictated by the WS dataflow. During the computation phase, four consecutive input values from the same row of matrix $A$ are fed into every TPE from the West side. To perform matrix multiplication, up to two of the four input elements are selected (using multiplexers) and multiplied with the locally-stored weights. The selection of the suitable input value(s) is facilitated by two 4-to-1 multiplexers that are controlled by the column indexes of the stored weights. The resultant products in each TPE are accumulated downwards across each column of the tensor array. When structured sparsity of $1$:$4$ is enabled, only one of the two multiplexers and multipliers will be activated in each TPE. This process iterates for all incoming rows of matrix $A$; different rows of $A$ arrive at each TPE in blocks of four consecutive elements, and the stationary weights dictate which ones are selected for computation.

\section{Fault Detection in Sparse Tensor Arrays}
\label{s:proposed}

The first step toward a fault-tolerant sparse tensor array is a fast and light-weight (in terms of both hardware overhead and the impact on application performance) fault-\textit{detection} mechanism that can trigger an appropriate reaction.

\begin{figure}
    \centering
    \includegraphics[width=0.7\columnwidth]{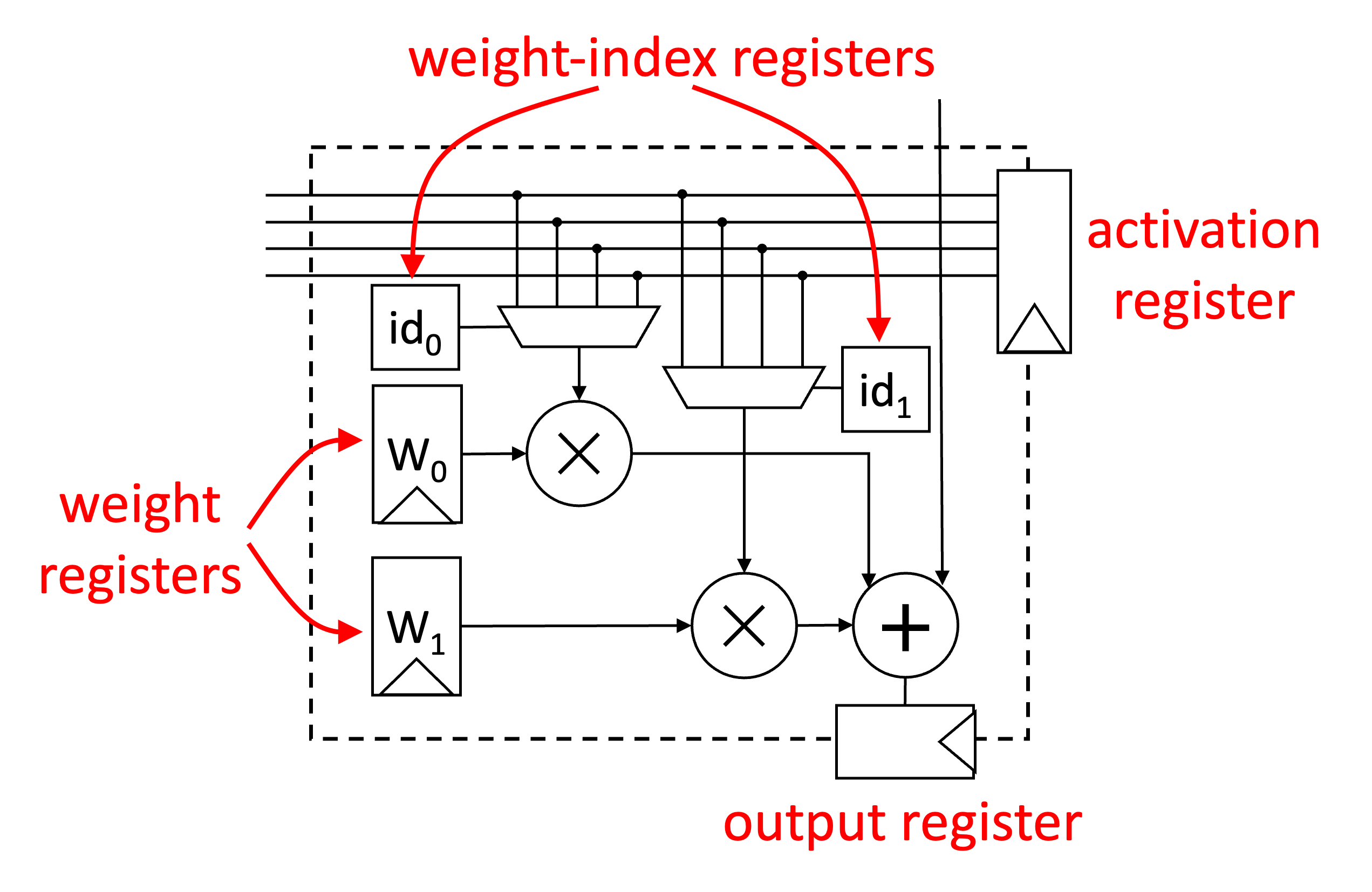}
    \caption{The four types of registers in a single TPE of a sparse systolic array.}
    \label{f:tpe_zoom}
\end{figure}

We hereby propose an online test technique that periodically checks -- every time a new group of weights is loaded -- the systolic array TPEs for permanent errors. For simplicity, let us assume that a permanent fault can only occur within the storage elements, i.e., the \textit{registers}, of each TPE, as shown in Fig.~\ref{f:tpe_zoom}. Hence, a permanent fault may afflict one of the following registers:

\begin{enumerate}
  \item The so called \textit{activation} registers, which store the incoming blocks of consecutive input elements and propagate them along the horizontal (west-to-east) direction in each row of the array.
  \item The $N$ \textit{weight} registers that store the stationary weights.
  \item The \textit{weight-index} registers that control the multiplexers and select the up to $N$ appropriate input (activation) elements out of the $M$-element input block.
  \item The \textit{output} registers that store the accumulated sums and propagate them downwards along the vertical direction (north-to-south) in each column of the array.
\end{enumerate}

Note that the simplifying assumption that permanent faults may only occur in the above-mentioned registers does \textit{not} ignore faults manifesting in the remaining logic of a PE. As can be seen in Fig.~\ref{f:tpe_zoom}, the micro-architecture of the TPE indicates that such faults will yield an erroneous result in either the multiplication and/or the accumulation steps, which will be captured as errors in the output register of each TPE. All paths of the internal TPE logic are ``funneled'' into the output register at the bottom right of each TPE. Hence, error coverage of the output register will also detect faults occurring within the remaining TPE logic. 

The proposed online checking mechanism detects and locates permanent (stuck-at) faults through a simple self-testing sequence. The sequence consists of four different tests, executed consecutively, with each test requiring a single test vector. Said test vectors are carefully crafted to reveal permanent faults occurring in any register within any TPE of the systolic array. The four test vectors use the weight values that are already stored (by the running application) in the array, thereby immensely reducing the test-latency overhead. The weights are not affected in any way and there is never a need to reload any data after a test session concludes. The application simply resumes normal operation.

The testing sequence is initiated after the weight-loading phase of each tile\footnote{Since the input and weight matrices of ML applications are typically larger than the size of the systolic array, the multiplication is performed gradually and progressively by loading \textit{tiles} of these matrices, whereby a tile corresponds to the size of the systolic array accelerator.} is completed. At this point, the pre-defined test vectors are fed as inputs to the systolic array and propagate through the array; i.e., they are multiplied-and-accumulated with the local weight values in each TPE along the way, until the final outputs are produced at the bottom of each column of the array (south edge).

\begin{figure}
    \centering
    \includegraphics[width=1 \columnwidth]{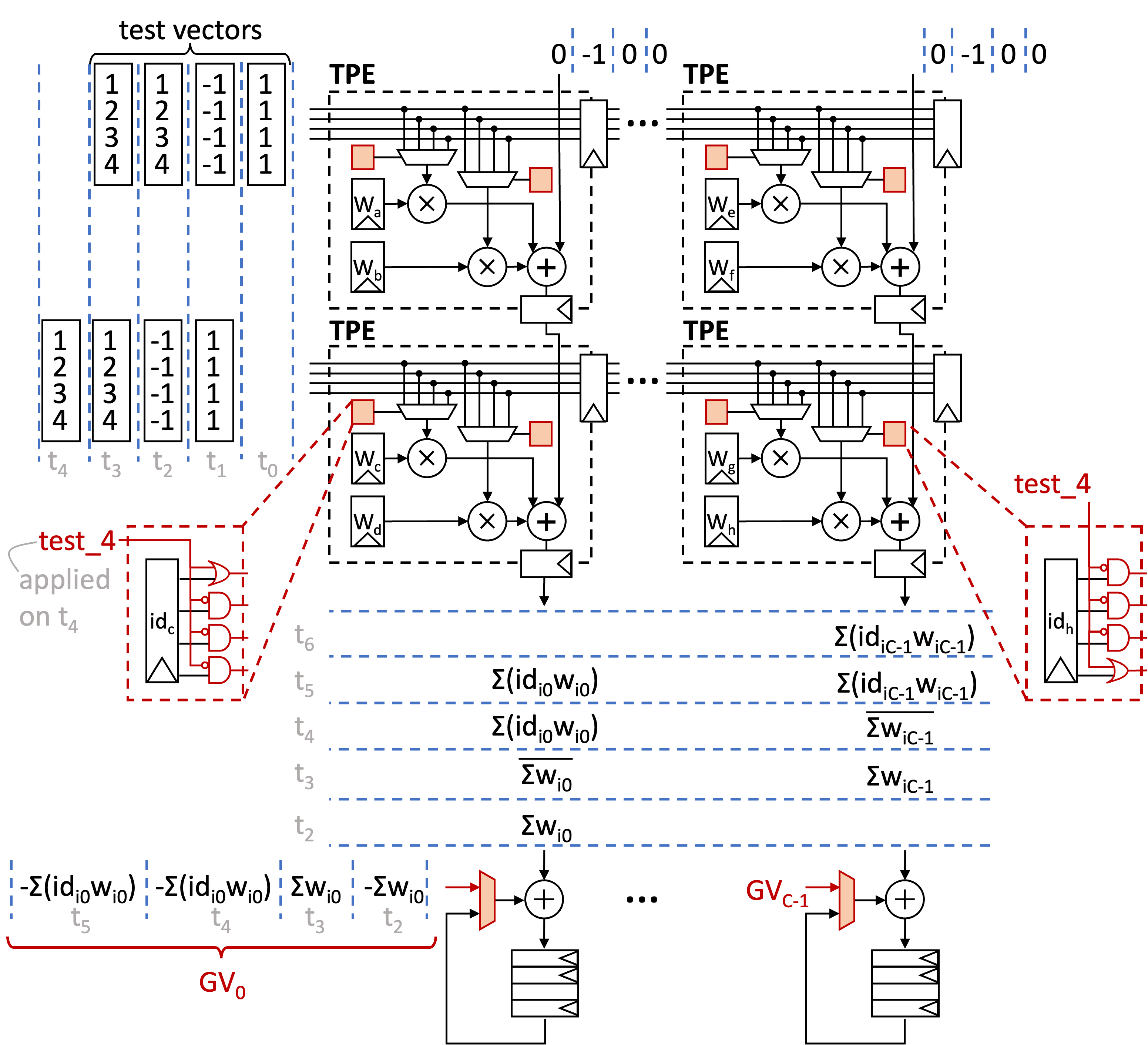}
    \caption{The four test vectors are applied periodically to the sparse systolic tensor array to detect permanent faults within any one register.}
    \label{f:test-vectors}
\end{figure}

A single testing session is depicted in Fig.~\ref{f:test-vectors} for the case that the inputs to each TPE are fed as groups of $M=4$ elements. Each test vector is fed into every row of TPEs in the array. For each test vector, there is an accompanying value that must be fed into the \textit{sum} (adder) input of the top-row TPEs of the array. The test vectors and the corresponding values fed into the top-row TPE adders are shown in Table~\ref{t:vectors}. The last two tests use the same test vector.

When fed into the array, the test vectors trigger the calculation of a ``weighted'' sum of all the weights already stored in the TPEs across each column of the array. This weighted sum is output at the bottom of each column and is then compared with the corresponding golden reference. This `comparison' is facilitated by the existing accumulators at the south edge of the SA, highlighted in red in Fig.~\ref{f:test-vectors}. In other words, the comparison is simply an addition operation.

Due to the use of the existing, specific weight values, it is not possible to ensure fault coverage for all possible paths in the design, i.e., achieve 100\% fault coverage. Nevertheless, if the test response is error free, there is very high certainty that, \textit{for the currently-loaded weight tile}, there will be no error affecting the output of the array. In other words, functional correctness is ensured with very high certainty \textit{for this particular weight tile}. The test must then be repeated when the next tile is loaded, and so forth. 

Also, the localization granularity is limited to a single column of the systolic array. In other words, the proposed mechanism can locate the faulty column of the array, but it cannot identify the specific TPE where the fault is located.

\begin{table}
    \centering
    \caption{The 3 unique test vectors employed by the proposed online checking methodology}
    \begin{tabular}{|c|c|c|}
        \hline
         Test number & Test vector & Top-row TPE sum input  \\ \hline \hline
              1      & $[1,1,1,1]$ &          $0$             \\
              2      & $[-1,-1,-1,-1]$ &     $-1$             \\
              3,4    & $[1,2,3,4]$ &          $0$            \\ \hline
    \end{tabular}
    \label{t:vectors}
\end{table}

\subsection{Testing for faults in the weight and output registers}

The first test vector computes the sum $V$ of the weight values across all TPEs in each column $j$, i.e., $V_{j}=\sum_{i} w_{ij}$. Index $i$ refers to all the weight registers across column $j$. For this first test, the golden reference value equals
\begin{equation*}
    GV_{j}=-\sum_{i} w_{ij}
\end{equation*}
\noindent The minus sign is included so that the \textit{addition} of the computed test sum $V_{j}$ with the golden reference value $GV_{j}$ should result in a zero value in the absence of a fault within each column $j$.

The second test vector -- used in the second test of each test session -- computes the \textit{bit-wise complement} of the sum of the weight values across all TPEs in each column $j$, i.e., $\overline{V_{j}}$. This is because the second test subtracts one (the input value fed into the top-row TPE) from the negative sum (due to the $-1$s in the test vector) of the weights of each column. 
By the 2's complement definition, the subtraction of $1$ from a negative signed operand effectively computes its bit-wise complement:
\begin{equation*}
    -V_{j}=\overline{V_{j}}+1 \Rightarrow  -V_{j}-1=\overline{V_{j}}  
\end{equation*}

Thus, when the bit-wise complement of the sum of weights, $\overline{V_{j}}$ is added to a golden reference value of 
\begin{equation*}
    GV_{j}=\sum_{i} w_{ij}\;\;\;(=\textrm{error-free}\;V_{j})
\end{equation*}
\noindent the result should be $-1$ (i.e., a bit string of all $1$s in the output value, since we use 2's complement arithmetic) in the absence of a fault, since $\overline{V_{j}}+V_{j}=-1$.

In summary, under fault-free operation, the results of the first test vector at the south edge of the array should be $0$ for all the columns, while the results of the second test vector should be $-1$ (all $1$s in the output values).

If there is any discrepancy in these first two tests, further checking is needed to locate the fault. As described in~\cite{runsafer} for dense systolic arrays, this fault localization is achieved by examining the outputs of Test 1 and Test 2 \textit{prior} and \textit{after} the comparison with the golden reference values, and checking whether they are bit-wise complementary, or not. Based on these bit-wise complementary checks, the fault can be localized using the cases depicted in Table~\ref{t:error-comparison}. Note that the localization identifies the register type that is faulty within a particular column of the array, but it \textit{cannot identify the specific TPE} in the column where the faulty register is located.

\begin{table}[h!]
    \centering
    \caption{Localization of faults by checking if the systolic array outputs and the comparison (with the golden reference) outputs of Tests 1 and 2 are bit-wise complementary}
    \label{t:error-comparison}
    \begin{tabular}{|l||p{1.7cm}|p{1.7cm}|p{1.7cm}|}
       \hline
       \textbf{Test 1 and 2} & \multirow{4}{*}{Complementary} & \multirow{2}{*}{Not} & \multirow{4}{*}{Complementary} \\
       \textbf{outputs before} &  &        & \\
       \textbf{comparison with} &    & \multirow{2}{*}{Complementary} & \\
       \textbf{golden reference} &    &  & \\\hline
       \textbf{Test 1 and 2}  & \multirow{4}{*}{Complementary} & \multirow{2}{*}{Not} & \multirow{2}{*}{Not}  \\
       \textbf{results after}  &  &  &   \\
       \textbf{comparison with}      &          & \multirow{2}{*}{Complementary} & \multirow{2}{*}{Complementary}  \\
       \textbf{golden reference}      &          &  &   \\\hline \hline
       \textbf{Fault}     & Weight & Output   & Comparison   \\
       \textbf{Location}  & register     & Register & Adder   \\\hline
    \end{tabular}
\end{table}

\subsection{Testing for faults in the weight-index registers}

The third test vector -- as used in the third test of each test session -- targets potential faults in the weight-index registers of each TPE in the array. The vector computes the ``weighted'' sum of the weight values, multiplied by their index position in the $M$-element block, as follows:
\begin{equation*}
    V_{j}=\sum_{i}(idx_{ij}\times w_{ij})    
\end{equation*}

\noindent Similar to the first test, the golden value reference is the negative of this sum, i.e., $GV_{j}=-V_j$
Consequently, if a fault occurs in any of the weight-index registers within a column of the array, a wrong element of the input test vector $[1,2,3,4]$ (assuming here that $M$=4) will be selected to be multiplied with a stationary weight, thereby resulting in an erroneous sum at the bottom of the column.

\subsection{Testing for faults in the input activation registers}

The first three test vectors target faults that affect the \textit{vertical} flow in the systolic array; i.e., faults that only affect the end result at the bottom of a column. The fourth test vector targets faults within the input activation registers of the \textit{horizontal} flow within the array (see Fig.~\ref{f:tpe_zoom}).

This test is more complex in sparse arrays than in dense arrays, since an erroneous activation value -- due to a fault within an activation register -- may not be selected for calculation until several columns later. Thus, we introduce another, more sophisticated test, as compared to~\cite{runsafer}, where a fault in any activation register will be `coerced' to manifest itself as multiple errors in consecutive columns.

This is achieved by utilizing a new control signal (labeled `test\_4' and shown in red in Fig.~\ref{f:test-vectors}) and masking gates at the outputs of the weight-index registers (also shown in red in Fig.~\ref{f:test-vectors}). When the new control signal is de-asserted, the masking gates are transparent, i.e., they let the output of the weight-index register pass through. The control signal is asserted only in the clock cycle that Test 4 is performed and it `activates' the masking gates. These force the multiplexers to select one specific value from the $M$-element input block. Specifically, in the first column, the first input element is selected (by the OR masking gate in the top-most position); in the second column, the second input element is selected (the OR gate is one position below), and so forth.

This will force the erroneous value to manifest at the output (south edge) of a column \textit{periodically}, every $M$ columns. That is, errors will be observed in \textit{multiple} columns, all spaced $M$ columns apart. Such behavior indicates that the fault occurred in an activation register in one of the $M$ columns located just before (to the left of) the first error appearance.

\section{Experimental Results}
\label{s:eval}

The experimental evaluation aims to quantitatively assess: (a) the fault-detection capability of the proposed online checking mechanism; (b) the impact on system performance in a fault-free environment; and (c) the hardware area cost. 

An 8$\times$8 sparse systolic tensor array was fully implemented in SystemVerilog RTL and augmented with the proposed checking mechanism. This systolic array operates on 16-bit integer quantized inputs and weights executing single-batch inference of CNNs that require matrix multiplications of different sizes. To retain accuracy, the additions in each column of the array are performed at a 32-bit width. The array was synthesized with Cadence's digital implementation flow using a 28 nm standard-cell library. It operates at a clock frequency of 1 GHz.

Three well-known CNNs were used for all the experiments: ResNet50~\cite{resnet}, DenseNet121~\cite{densenet} and VGG16~\cite{vgg}.

Since the focus of this evaluation is on fault-detection, the experiments were conducted at the synthesized gate-level netlist. To calculate the achieved fault coverage, the Hope sequential fault-simulator~\cite{hope} was employed, which performs exhaustive fault simulation at the gate-level of the entire systolic array. Specifically, Hope performs a single-fault injection campaign for each point in the given netlist and checks if the application of the applied test-vectors (in our case: the 4 test vectors and all the weight values in the CNN application) produce an output different from the fault-free output. 

\begin{figure}
    \centering
    \includegraphics[width=0.9\columnwidth]{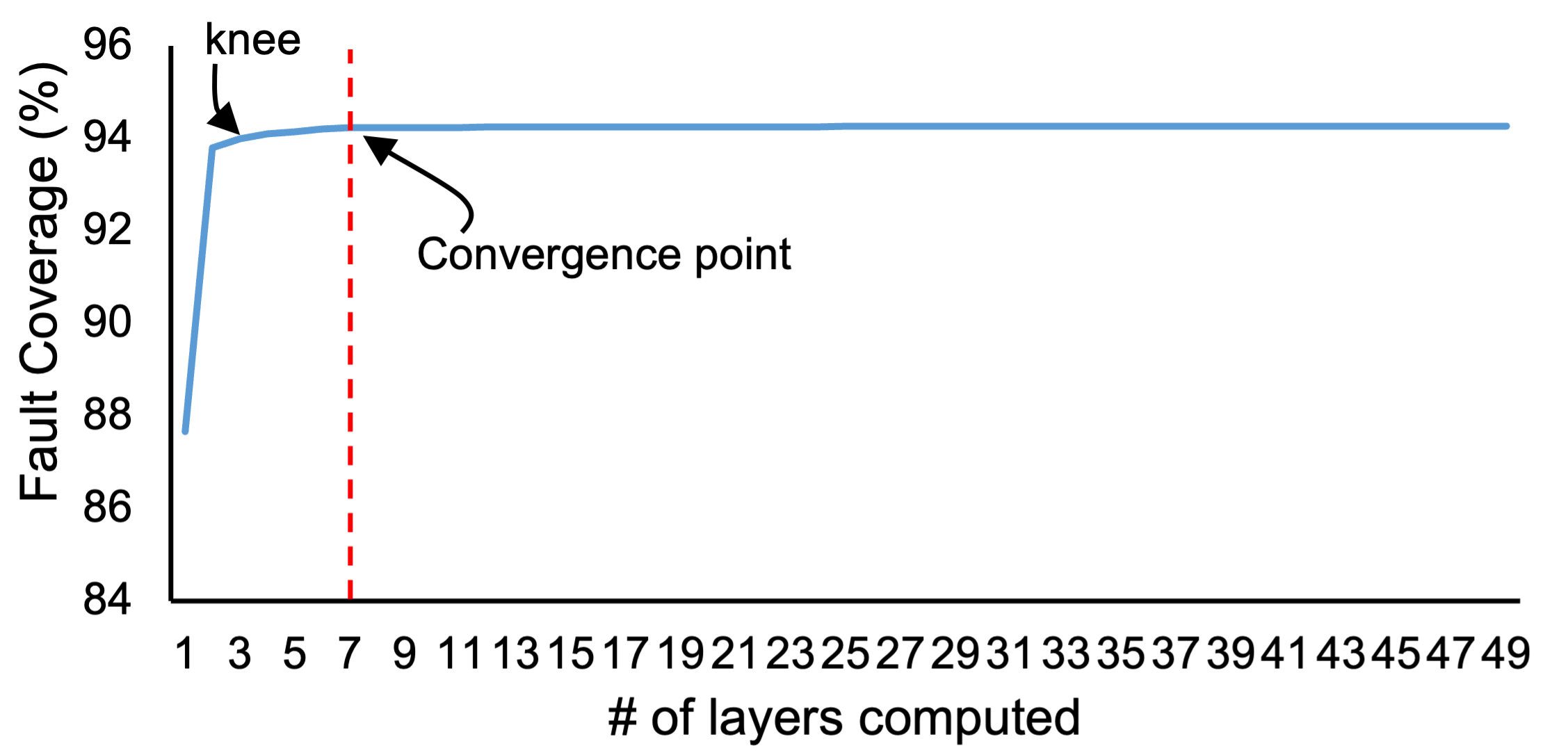}
    \caption{The fault coverage achieved after completion of each layer of ResNet50~\cite{resnet}. The fault coverage converges quite rapidly to a high value.}
    \label{f:resnet-per-layer-cov}
\end{figure}

When a new tile of weights is loaded, four distinct tests are performed in consecutive clock cycles using the four proposed test vectors applied as inputs to the array. Consequently, the achieved fault coverage increases progressively as more and more weight tiles are loaded during the execution of the CNN application. Fig.~\ref{f:resnet-per-layer-cov} depicts this increase in the achieved fault coverage as all the layers of ResNet50~\cite{resnet} pass through the systolic array and a testing session is performed on each new weight tile. After the first CNN layer is completed, the fault coverage is quite low, at 88.7\%. However, the fault coverage increases rapidly within the first three layers. At the end of the third layer, there is a `knee,' beyond which the increase in the fault coverage is minimal. A convergence point is reached after completion of the seventh layer, after which the changes in fault coverage are negligible.

Table~\ref{t:fault-coverage} summarizes the achieved fault coverages and convergence points for all three examined CNN applications. As shown, all three CNN applications converge quite rapidly (after only a few layers) to an average fault coverage of 94.2\%.

\begin{table}
    \centering
    \caption{Fault coverage achieved by the proposed online testing mechanism for three well-known CNN applications and the layer of convergence for each application.}
    \begin{tabular}{|l|c|c|c|}
        \hline
         \multirow{2}{*}{\textbf{App}}& \multirow{2}{*}{\textbf{Total \# of layers}} & \multirow{2}{*}{\textbf{Fault Coverage}} & \textbf{Convergence}  \\
                     &                             &                         & \textbf{layer}  \\\hline \hline
         ResNet50    & 49  &     94.2\%              &  7  \\ 
         DenseNet121 & 120 &     94.3\%              &  2  \\
         VGG16       & 13  &     94.1\%              &  5  \\
         \hline
    \end{tabular}
    \label{t:fault-coverage}
\end{table}

Obviously, since the proposed mechanism uses only 4 test vectors and the existing weights in each loaded weight tile, the achieved gate-level stuck-at fault coverage cannot reach 100\%; the weights are not sufficient to exercise all stuck-at faults. Our observations indicate that the vast majority of uncovered faults are paths located inside the multipliers of each TPE, which are not activated due to the constant weight values. However, as previously mentioned, even if the \textit{gate-level} fault coverage is not closer to 100\%, an error-free test response indicates with very high certainty \textit{functional} correctness \textit{for the currently-loaded weight tile}; i.e., there will be no error affecting the output of the array during the execution of this particular weight tile.

\begin{figure}
    \centering
    \includegraphics[width=0.7\columnwidth]{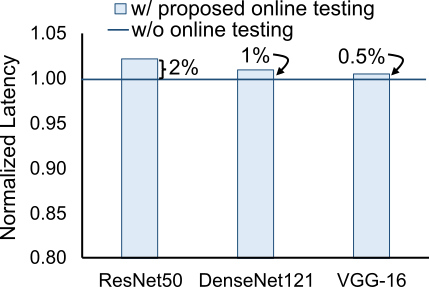}
    \caption{The impact on application latency of the proposed periodic online testing methodology. The results are normalized to the latency of the sparse systolic array with no error-checking mechanism.}
    \label{f:norm-latency}
\end{figure}

To evaluate the latency overhead introduced by the proposed periodic online testing mechanism, the total runtime of each CNN application \textit{in the presence of online testing} was measured. An online test session was triggered for each newly-loaded weight tile throughout the entire execution of the application. This measured runtime was compared to the runtime achieved \textit{without any online testing}, i.e., the execution latency achieved on a baseline sparse systolic tensor array. The results for all three CNN applications are shown in Fig.~\ref{f:norm-latency}. The proposed checking mechanism adds only marginal latency overhead of $0.5\%$ -- $2\%$ to the applications' total runtimes. This is due to the use of only 4 test vectors, which merely add a 4-cycle latency overhead for every executed weight tile.

The hardware area overhead to support the proposed mechanism is also minimal. As shown in red in Fig.~\ref{f:sparse-tensor-array}, the extra hardware added comprises the single `test\_4' control signal, three AND gates and one OR gate per weight-index register in each TPE, and a multiplexer at the input of the bottom accumulators of each column, at the south edge of the SA. The total additional hardware accounts for 3\% of the total area of the sparse systolic tensor array.

\section{Related Work}
\label{s:related-work}

There are numerous efforts aimed at protecting systolic array architectures from faults, which can be categorized into three main approaches: (a) detection-only methods that identify faults and discard faulty results; (b) detection and localization methods that isolate erroneous hardware modules after detecting faults; and (c) detection, localization, and correction methods that also attempt to correct faults for faster system recovery. These fault-tolerant methodologies are applied to dense systolic arrays and vary in complexity and effectiveness.

For fault detection, the work in~\cite{access} uses connections between weight and activation registers to form new scan-chains, reducing hardware complexity and power consumption. Another method~\cite{low-voltage-sa} employs ABFT to detect faults caused by voltage reductions in low-power systolic arrays, while~\cite{online-date23} uses extra accumulators to implement ABFT directly on Intel's Tiled Matrix Multiplication (TMUL) units. In a similar vein,~\cite{aicas-error-checking} adapts the ABFT methodology to the unique characteristics of sparse systolic tensor arrays and to Graph Convolutional Networks (GCN)~\cite{GCN-abft}. These methods primarily focus on identifying faults without correcting them, thereby preventing faulty results from impacting system operations.

More advanced methodologies not only detect, but also localize and correct faults. Techniques like those in~\cite{zhang2018analyzing} and~\cite{sadi2021test} combine fault-aware pruning and retraining to minimize accuracy degradation and to bypass critical faults in Tensor Processing Units (TPUs). Additionally, the work in~\cite{lee2023new} tests near-threshold systolic array architectures without extra hardware, and RunSafer~\cite{runsafer} uses specific test vectors to protect against permanent faults with minimal latency overhead. Correction methods like STRAIT~\cite{strait} use self-test and recovery architectures to address faults, employing weight pruning and row/column swapping to maintain accuracy. Another approach, in~\cite{ma2023error}, focuses on error detection and correction in Transformer networks, mitigating out-of-range neuron outputs through saturation or zeroing, thus ensuring minimal accuracy degradation even under high error rates. Finally, the work in~\cite{agarwal2023towards} introduces fault-injection frameworks to analyze the impact of various factors on fault propagation in systolic arrays, further enhancing fault-tolerance strategies.

\section{Conclusions and Future Work}
\label{s:conclusions}

This work addresses the need for fault-tolerant ML accelerator hardware in safety-critical applications like autonomous vehicles, medicine, and aviation. The proposed periodic online testing targets \textit{sparse} systolic tensor arrays used for structured-sparse ML models. Each test session requires only four test vectors, the model’s existing weights, and simple comparisons with precomputed reference values to detect permanent faults. Unlike ABFT, testing occurs \textit{before} execution, preventing wasted work. Moreover, the location of any detected fault is also identified at column-level granularity. The experimental evaluation demonstrates high stuck-at fault coverage with minimal impact on application performance and minimal additional hardware. As a direction for future work, we plan to enhance the current test-vector set by incorporating a small number of strategically selected random vectors. These will be designed to exercise previously unreachable paths, thereby further improving overall fault coverage without significantly increasing testing overhead.

\bibliographystyle{IEEEtran}
\bibliography{refs}

\end{document}